\documentclass[english]{article}
\usepackage[T1]{fontenc}
\usepackage[latin9]{inputenc}
\usepackage[letterpaper]{geometry}
\usepackage{color}
\usepackage{array}
\usepackage{amsmath}
\usepackage{graphicx}

\makeatletter

\providecommand{\tabularnewline}{\\}

\usepackage{array}
\usepackage{textcomp}
\usepackage{fix-cm}

\providecommand{\tabularnewline}{\\}

\usepackage{babel}

\usepackage{babel}

\usepackage{esint}
\usepackage{epstopdf}
\usepackage{epsfig}

\makeatother

\usepackage{babel}
\begin{document}

\title{Asymmetric Quantum Dialogue in Noisy Environment}

\author{Anindita Banerjee$^{a,}$\thanks{email: anindita.phd@gmail.com},
Chitra Shukla$^{b,}$\thanks{email: shukla.chitra@i.mbox.nagoya-u.ac.jp },
Kishore Thapliyal$^{c,}$\thanks{email: tkishore36@yahoo.com}, Anirban
Pathak$^{c,}$\thanks{email: anirban.pathak@jiit.ac.in}, Prasanta
K. Panigrahi$^{d,}$\thanks{email: pprasanta@iiserkol.ac.in}}

\maketitle
\begin{center}
$^{a}$Department of Physics and Center for Astroparticle Physics
and Space Science, Bose Institute, Block EN, Sector V, Kolkata 700091,
India 
\par\end{center}

\begin{center}
$^{b}$Graduate School of Information Science, Nagoya University,
Furo-cho 1, Chikusa-ku, Nagoya, 464-8601, Japan
\par\end{center}

\begin{center}
$^{c}$Jaypee Institute of Information Technology, A-10, Sector-62,
Noida, UP-201307, India
\par\end{center}

\begin{center}
$^{d}$Department of Physical Sciences, Indian Institute of Science
Education and Research Kolkata, Mohanpur 741246, India
\par\end{center}
\begin{abstract}
A notion of asymmetric quantum dialogue (AQD) is introduced. Conventional
protocols of quantum dialogue are essentially symmetric as both the
users (Alice and Bob) can encode the same amount of classical information.
In contrast, the scheme for AQD introduced here provides different
amount of communication powers to Alice and Bob. The proposed scheme,
offers an architecture, where the entangled state and the encoding
scheme to be shared between Alice and Bob depends on the amount of
classical information they want to exchange with each other. The general
structure for the AQD scheme has been obtained using a group theoretic
structure of the operators introduced in (Shukla et al., Phys. Lett.
A, \textbf{377} (2013) 518). The effect of different types of noises
(e.g., amplitude damping and phase damping noise) on the proposed
scheme is investigated, and it is shown that the proposed AQD is robust
and uses optimized amount of quantum resources.
\end{abstract}
\textbf{Keywords:} Asymmetric quantum dialogue, noise models, secure
quantum communication.

\section{Introduction\label{sec:Introduction}}

The birth of quantum cryptography in 1984 \cite{bb84}, followed by
its variants steadily paved way for various quantum communication
schemes, thereby, establishing its manifold applications (see \cite{book}
and references therein). Initial protocols \cite{bb84,ekert,b92,vaidman-goldenberg}
of secure quantum communication were restricted to the designing of
protocols for quantum key distribution (QKD), which enables a sender
(Alice) to share (distribute) a key with a receiver (Bob). It was
later realized that a pre-shared (pre-distributed) key is not essential
for secure quantum communication, and several schemes for secure direct
quantum communication were proposed \cite{ping-pong,dsqc-1,review,Man,Anindita,cdsqc}.
These schemes of secure direct quantum communication were mainly of
two types: quantum secure direct communication (QSDC) and deterministic
secure quantum communication (DSQC). In a QSDC scheme, the receiver
can decode an information encoded by the sender without any additional
information (classical communication) \cite{ping-pong,dsqc-1,review}.
In contrast, in a scheme for DSQC, the receiver requires at least
one bit of classical information to decode each bit of the message
encoded by the sender \cite{Man,Anindita}. Different variants of
DSQC and QSDC protocols have been studied in the recent past. Specifically,
controlled DSQC has been studied in much detail in the recent past
(\cite{cdsqc} and references therein). Interestingly, the schemes
of DSQC and QSDC are one way schemes in the sense that these schemes
only allow Alice to communicate a message to Bob, but does not allow
Bob to do the same. In contrast, in daily life, one often need bidirectional
communication, where two parties can simultaneously communicate messages
to each other, an example been classical communication via telephone.
Keeping this in mind, a scheme for simultaneous bidirectional quantum
communication was introduced by Ba An \cite{ba-an} in 2004, and was
referred to as ``quantum dialogue (QD)''. However, it was found
that the Ba An scheme was insecure under intercept-resend attack \cite{Man}.
In Ref. \cite{Man}, an effort was made to modify the original Ba
An scheme to provide a secure scheme for QD. However the effort was
not successful and later a secure version of the original scheme was
proposed by Ba An himself \cite{baan_new} using Bell states.

It is important to note that a scheme for QD is different from a system
where conversation between Alice and Bob happens using two QSDCs or
two DSQCs (say, one QSDC/DSQC is used for Alice to Bob communication
and the other one is used for Bob to Alice communication). In QD,
messages of Alice and Bob are required to be simultaneously encoded
in the same channel. Further, we would like to note that this type
of scheme for bidirectional direct quantum communication has been
referred by various names, but schemes proposed under these alternate
names are equivalent and may be referred to as quantum dialogue. Specifically,
equivalent bidirectional schemes for secure direct quantum communication
have been referred to as bidirectional quantum communication in \cite{shi-auxilary},
quantum telephone in \cite{quantum telephon1,Y Sun improve telephone},
quantum dialogue in \cite{ba-an,Naseri}, quantum conversation in
\cite{sakshi-panigrahi-epl}, etc. 

The practical importance of the QD drew considerable attention of
the quantum cryptography community and several schemes for QD have
been proposed. For example, Xia et al. proposed a QD scheme using
$GHZ$ states \cite{xia} and Dong \emph{et al.} proposed the same
using tripartite $W$ states \cite{dong-w}. Further, efforts have
been made to realize QD using various approaches. For instance, protocols
for QD have been proposed using (i) dense coding \cite{Man,quantum telephon1,xia},
(ii) entanglement swapping \cite{gao-swapping,QD-EnSwap}, (iii) single
photon \cite{Naseri}, (iv) auxiliary particles \cite{shi-auxilary},
(v) qutrit states \cite{QD-EnSwap,QD-qutrit,com-QD-qut}, (vi) continuous
variable states \cite{CV-QD}, (vii) prior shared key for authentication
\cite{Naseri,probAuthQD}, etc. Apart from all these variants of QD,
a modified QD scheme with the name quantum secure direct dialogue
\cite{QSDD1,QSDD2} has been recently proposed, in which, a few of
the Bell states are used for Alice to Bob communication, while the
remaining for sending Bob's message to Alice. This is not really a
scheme for QD as the information encoded by Alice and Bob does not
exist simultaneously in the same channel. Thus, it is to be viewed
as two QSDCs. In the scheme proposed in \cite{QSDD1,QSDD2}, Alice
and Bob were capable of sending messages of different length to each
other, and this was claimed to be the advantage of quantum secure
direct dialogue protocol \cite{QSDD1,QSDD2} over conventional protocols
of QD. However, to achieve this feature they had to go beyond the
domain of QD. In what follows, we will show that we can achieve this
feature, strictly remaining within the domain of quantum dialogue.
Further, we would like to note that almost all the Ba-An-type QD schemes
proposed so far have a generalized mathematical structure \cite{qd}.
Specifically, if a maximally densecodable quantum state is used as
quantum channel, the encoding operations form a group under multiplication.
This point will be further discussed in detail in the forthcoming
sections.

Recently, considerable efforts have also been made to avoid information
leakage in quantum dialogue protocols (see \cite{Y Sun improve telephone,gao-swapping,QD-EnSwap,com-QD-qut}
and reference therein). Some efforts have also been made to study
the effect of noise on the schemes of QD \cite{vishal,crypt-switch,QD-hwang}.
For instance, in Ref. \cite{vishal}, the effect of a set of noise
models on a single-particle-based and an entangled-state-based QD
schemes has been investigated for a comparative analysis. Similarly,
the effect of noise on the controlled version of a QD scheme is analyzed
under amplitude and phase damping noise models in Ref. \cite{crypt-switch}.
Further, Yang and Hwang \cite{QD-hwang} have proposed a QD scheme
immune to collective noise using a set of logical bits (which are
Bell states immune to collective noise) as reported in \cite{col-Bell}.
Specifically, some Bell states form a decoherence free subspace under
collective noise and have been used to design various QD protocols
since then \cite{QSDD2,QD-hwang,QD-GHZ}. However, all the schemes
of QD proposed so far and investigated under noisy environment are
symmetric in the sense that Alice and Bob encode equal amount of information.
Interestingly, in practical circumstances, one often comes across
situations, where one of the users communicates more (say in an online
lecture class usually Professor speaks more). In such situations,
we need schemes for asymmetric quantum dialogue (AQD), where the communication
power (the amount of classical information one can send) is different
for different users. In what follows, we have referred to standard
schemes of symmetric QD as QD and have designed a new scheme of AQD,
which is shown to be interesting by establishing that it would require
lesser quantum resource and be less affected in a noisy channel (amplitude
damping and phase damping channel). 

The remaining part of this paper is organized as follows. In Section
\ref{sec:Protocol-for-asymmetric}, the protocol of AQD is proposed,
and subsequently, the group theoretic structure of the operators that
can be used to implement the protocol is discussed in Section \ref{sec:Group-theoretic-structure}.
In Section \ref{sec:Robustness:-Effect-of}, the feasibility of the
proposed AQD scheme is analyzed under the effect of amplitude damping
and phase damping noise. The leakage and efficiency in the proposed
AQD scheme are discussed in Section \ref{sec:Leakage} and finally,
the paper is concluded in Section \ref{sec:Conclusion}.

\section{Protocol for asymmetric quantum dialogue \label{sec:Protocol-for-asymmetric}}

Before we proceed with the AQD protocol, it would be appropriate to
note that the protocol proposed in this section is a Ba-An-type protocol,
and is along the line of the generalized Ba-An-type protocol for QD
proposed by some of the present authors in Ref. \cite{qd}, which
we refer to as Shukla-Banerjee-Pathak (SBP) protocol. In SBP protocol,
an $n$-qubit quantum state $|\phi_{1}\rangle$ from the basis set
$\left\{ |\phi_{i}\rangle\right\} $ is used. Further, a set of $2^{n}$
unitary operators $\{U_{1},U_{2},\cdots,U_{2^{n}}\}$ are needed to
encode an $n$-bit message, where the unitary operators are essentially
$j$-qubit ($j\leq n$) operators such that $U_{i}|\phi_{1}\rangle=|\phi_{i}\rangle$.
It is also necessary that the set of operators $\{U_{1},U_{2},\cdots,U_{2^{n}}\}$
forms a group under multiplication after neglecting the global phase
\cite{qd}. This is important because in Ba-An-type schemes, Alice
encodes $U_{A}$ on the quantum state $|\phi_{1}^{\prime}\rangle=U_{B}|\phi_{1}\rangle$
which was produced by Bob by applying a  unitary operation $U_{B}$
on the initial state $|\phi_{1}\rangle$. Thus, the final state after
Alice's operation becomes $|\phi_{1}^{\prime\prime}\rangle=U_{A}U_{B}|\phi_{1}\rangle,$
which will be in the basis set $\left\{ |\phi_{i}\rangle\right\} $
only if $U_{A}U_{B}=U_{j}$ is an element of the set $\{U_{1},U_{2},\cdots,U_{2^{n}}\}$.
Its relevance would be clear if we note that, to decode any information
sent by Alice to Bob (or equivalently from Bob to Alice) one needs
knowledge of either $U_{A}$ or \textbf{$U_{B}$}. As only Alice and
Bob possess this information, they can extract the information encoded
by the other user if one of the user (Bob) measures $|\phi_{1}^{\prime\prime}\rangle$
using $\left\{ |\phi_{i}\rangle\right\} $ basis set and announces
the result. These conditions led to a generalized group theoretic
structure for the operators to be used to implement a protocol of
QD. In what follows, we use the same conditions and notations to introduce
a protocol for asymmetric quantum dialogue, where Alice and Bob wish
to send messages of $m$ and $n$ bits ($m\neq n$ for AQD), respectively
to each other. In contrast, in a conventional symmetric QD, one always
had $m=n$. Technically, it is possible to achieve the task accomplished
by AQD using a conventional scheme of QD, but that would require utilization
of more quantum resources. To be precise, if $m<n$ then Alice would
require to send $\left(n-m\right)$ auxiliary bits. Similarly, Bob
would require to send $\left(m-n\right)$ auxiliary bits, when $m>n$.
Without loss of generality, here we may describe our protocol for
AQD for the specific case $m<n$ and show that neither Alice needs
to send any auxiliary bits, nor we need to go beyond the domain of
quantum dialogue (as was done in Refs. \cite{QSDD1,QSDD2}). The protocol
for AQD can be described as follows: 
\begin{description}
\item [{AQD1}] Bob prepares $p$ copies of an initial state $|\phi_{1}\rangle,$
which is an $n$-qubit entangled state, i.e., he prepares $|\phi_{1}\rangle^{\otimes p}$.
Subsequently, he encodes his message by applying a $j$-qubit unitary
operator from the group of unitary operators $\{U_{1},U_{2},\cdots,U_{2^{n}}\}.$
\\
Note that each unitary operation encodes an $n$-bit classical message.
Hence, Bob encodes an $np$-bit message using all the states. The
orthogonality of the information encoded states is ensured due to
the specific properties of the set of unitary operators used here,
i.e., operation of an operator from the set $\{U_{1},U_{2},\cdots,U_{2^{n}}\}$
gives $U_{i}|\phi_{1}\rangle=|\phi_{i}\rangle,$ which is an element
of the complete basis set $\left\{ |\phi_{i}\rangle\right\} $.
\item [{AQD2}] Bob prepares two strings of travel and home qubits as $P_{A}=[p_{1}(t_{1},t_{2},\cdots,t_{l}),$
$p_{2}(t_{1},t_{2},\cdots,t_{l}),$$...,p_{N}(t_{1},t_{2},\cdots,t_{l})]$
and $P_{B}=[p_{1}(h_{1},h_{2},...,h_{n-l}),$ $p_{2}(h_{1},h_{2},...,h_{n-l}),$$\cdots,p_{N}(h_{1},h_{2},...,h_{n-l})]$,
respectively. The string of the travel qubits contains all the $l$
qubits on which Alice will encode her message in \textbf{AQD4}. Another
string of home qubits is composed of the qubits not encoded by Bob
and the Bob's encoded qubits on which Alice is not supposed to encode
her secret. Then Bob prepares $lp$ decoy qubits and concatenate them
with $P_{A}$ to form a larger string $P_{A}^{\prime}$. Subsequently,
he applies a permutation operator $\Pi_{2lp}$ on the string $P_{A}^{\prime}$
to obtain a new string $P_{A}^{\prime\prime}$ and sends it to Alice.
\\
A detailed discussion of the various types of decoy qubit based eavesdropping
check subroutines can be found in the recent literature (\cite{Kishore-decoy}
and references therein).
\item [{AQD3}] Bob announces the permutation operator $\Pi_{lp}$ corresponding
to the decoy qubits, i.e., the correct positions of the decoy qubits
on which certain decoy qubit based eavesdropping checking technique
(such as BB84 subroutine, GV subroutine, etc. \cite{book,Anindita,cdsqc,qd,QKA-cs})
depending upon the choice of decoy qubits can be applied \cite{Kishore-decoy}.
They proceed with the protocol if the detected error rate is found
to be below a tolerable limit, otherwise they start afresh. 
\item [{AQD4}] Bob informs Alice the order of the remaining qubits. Then
Alice obtains the actual order and performs her encoding using a set
of $l$-qubit unitary operators $\{U_{1}^{\prime},U_{2}^{\prime},\cdots,U_{2^{m}}^{\prime}\}$
such that $\{U_{1}^{\prime}\otimes I_{2}^{\otimes j-l},U_{2}^{\prime}\otimes I_{2}^{\otimes j-l},\cdots,U_{2^{m}}^{\prime}\otimes I_{2}^{\otimes j-l}\}$
forms a subgroup (of order $2^{m}$) of the group $\{U_{1},U_{2},\cdots,U_{2^{n}}\}$,
which contains the operators used by Bob for encoding his message,
and is a group of order $2^{n}$ with $n>m$ for AQD. Thereafter,
Alice prepares $lp$ decoy qubits and concatenates them with the original
string $P_{AB}$ to obtain a larger sequence $P_{AB}^{\prime}$. This
is followed by a permutation operation on the enlarged sequence $P_{AB}^{\prime}$
by Alice to create $P_{AB}^{\prime\prime}$. Finally, she sends $P_{AB}^{\prime\prime}$
back to Bob.\\
Note that the Identity operator $(I_{2})$ mentioned above, were included
in the description to illustrate the nature of the subgroup that can
be formed and in our case, we can visualize it as a situation in which
Alice encoded her message on the qubits received by her using an operator
$U_{i}^{\prime}$ from the group of operators $\{U_{i}^{\prime}\}$
and thus, $I_{2}$ operators operate on the remaining qubits on which
Bob had encoded his message, but not send to Alice. 
\item [{AQD5}] Bob also performs an eavesdropping checking (in collaboration
with Alice) as in \textbf{AQD3}. They proceed with the protocol if
and only if sufficiently low error rate is obtained, otherwise they
restart from \textbf{ADQ1}.
\item [{AQD6}] Alice announces the permutation operator for Bob to reorder
the remaining qubits. Bob obtains $P_{AB}$ and recombines it with
$P_{B}$ to measure each $n$-qubit entangled state in $\left\{ |\phi_{i}\rangle\right\} $
basis. Subsequently, Bob publicly announces the final states he had
obtained on measurement. The initial and final states are publicly
known. Bob knows his encoding as well, with the help of which he can
decode Alice's message. Similarly, Alice obtains Bob's message using
these publicly known information and her knowledge about the encoding
operation performed by her during \textbf{AQD4}. 
\end{description}
Specific examples of the AQD scheme will be discussed in the forthcoming
sections.

\section{Group theoretic structure of the AQD protocol \label{sec:Group-theoretic-structure}}

To illustrate the general structure of the possible schemes for AQD,
we may use a notion of the modified Pauli group introduced in \cite{qd,QKA-cs}.
In \cite{qd}, an operational definition of the Pauli group was used,
where global phase was ignored from the group multiplication table
of the Pauli group (thus, more than one element of standard Pauli
group \cite{nielsen} which are different only in global phase would
become a single element of the modified Pauli group \cite{qd}). For
the convenience of the reader, the operational definition of the modified
Pauli group and the notation introduced in \cite{qd} and followed
in this paper are summarized in Appendix A. In what follows, we describe
the group theoretic structure of the operators used by Alice and Bob
to realize the proposed scheme of AQD. As the choice of the encoding
operators depends on the entangled states to be shared between Alice
and Bob, we restrict our discussion to a finite set of $n$-qubit
entangled states (where we choose $2\leq n\leq5)$. Specifically,
for $n=2,$ only entangled state of interest is a Bell state; whereas
for $n=3,$ all arbitrary entangled state can be classified into two
sub-classes: $GHZ$-type states and $W$-type states, and keeping
that in mind for $n=3,$ we have restricted our investigation to the
search of groups of unitary operators that can be used to implement
QD or AQD using representative quantum states from $GHZ$ class and
$W$ class. Similarly, for $n=4$ case, we have concentrated on the
representative quantum states from the 9 families of 4-qubit entangled
states introduced in \cite{family}. We have already noted that all
3-qubit entangled states can be classified as $GHZ$-type states and
$W$-type states. Inspired by this observation Verstraete et al. tried
to classify 4-qubit entangled states into a finite set of SLOCC nonequivalent
families and introduced 9 families which were referred to as $G_{abcd},L_{abc_{2}},L_{a_{2}b_{2}},L_{ab_{3}},L_{a_{4}},L_{a_{2}0_{3\oplus\bar{1}}},L_{0_{5\oplus\bar{3}}},L_{0_{7\oplus\bar{1}}},L_{0_{3\oplus\bar{1}}0_{3\oplus\bar{1}}}$
(for exact definitions of these families see Theorem 2 of Ref. \cite{family}).
Later on, Chterental et al. \cite{4-Qubit-SLOCC-proof} and Borsten
et al. \cite{string} also obtained these nine families through different
approaches. However, Gour and Wallach had later shown that actually
there exist an infinite number of SLOCC non-equivalent classes for
four qubit entangled states \cite{4qubit-inf}. It's not our purpose
to discuss these SLOCC nonequivalent classes in detail. Rather, we
are interested in illustrating the group theoretic structure of the
AQD scheme with some quantum states as an example, and for this purpose
we have chosen representative states from different families of 4-qubit
entangled states as classified in \cite{family}. Finally, we also
report the group theoretic structure of the operators that may used
to implement QD/AQD using specific type of 5-quabit entangled states
(namely 5-qubit Brown and cluster states).

To begin with, we note that a large number of alternate possibilities
of implementing QD using $n$-qubit ($2\leq n\leq5$) entangled states
and various groups of unitary operators were already listed in Table
4 of our earlier work \cite{qd}. The present investigation has revealed
a number of new possibilities, and they are summarized in Table \ref{table_group operators}.
Specifically, the last column of Table \ref{table_group operators}
reports a new set (i.e., not reported earlier) of group of operators
that may be used to implement QD or AQD using a particular type of
entangled state mentioned in the first column of the table. Table
\ref{table_group operators} clearly illustrates that a scheme for
QD can be implemented using a state mentioned in the 1st column of
the $i$th row of the table, if the users use one of the groups listed
in Column 3 or 4 of the same row. However, to implement an AQD, using
a state mentioned in the 1st column of $i$th row of this table, one
of the user (say, Bob, who is the first user in the sense that he
prepares the quantum channel) has to use a group (say, $G_{B}$) of
order $2^{n}$ listed in Column 3 or 4 of the same row, whereas the
other user (Alice) would require to use another group of operators
(say, $G_{A}$) of order $2^{m},$ such that $G_{A}\otimes\{I_{2}^{j-l}\}$
(cf. \textbf{AQD4}) forms a subgroup of $G_{B}$. In what follows,
we discuss a few specific examples to extend this point, and provided
a large list of allowed combination of states and such groups of operators
in Table \ref{tab:Assym-QD}.

\begin{table}
\centering{}{\footnotesize{}}%
\begin{tabular}{|c|>{\centering}p{0.5in}|>{\centering}p{2.4in}|>{\centering}p{2.1in}|}
\hline 
{\footnotesize{}Quantum state } & {\footnotesize{}SLOCC nonequivalent family} & {\footnotesize{}Group of unitary operations that can be used for QD
and described in Ref. \cite{qd}} & {\footnotesize{}New group of unitary operations that can also be used
for QD }\tabularnewline
\hline 
{\footnotesize{}2-qubit Bell state } & Bell & {\footnotesize{}$G_{1}$} & \tabularnewline
\hline 
{\footnotesize{}3-qubit $GHZ$ } & $GHZ$ & {\footnotesize{}$G_{2}^{1}(8),G_{2}^{2}(8),G_{2}^{4}(8),G_{2}^{5}(8)$} & \tabularnewline
\hline 
{\footnotesize{}3-qubit $GHZ$-like } & $GHZ$ & {\footnotesize{}$G_{2}^{2}(8),G_{2}^{3}(8),G_{2}^{5}(8),G_{2}^{6}(8),G_{2}^{8}(8),G_{2}^{9}(8)$} & \tabularnewline
\hline 
{\footnotesize{}4-qubit cat state} & $G_{abcd}$ &  & {\footnotesize{}$G_{2}^{1}(8),G_{2}^{2}(8),G_{2}^{4}(8),G_{2}^{5}(8)$}\tabularnewline
\hline 
{\footnotesize{}4-qubit $W$ } & $L_{ab_{3}}$ & {\footnotesize{}$G_{2}^{8}(8),G_{2}^{9}(8)$} & \tabularnewline
\hline 
{\footnotesize{}4-qubit $Q_{5}$ } & {\footnotesize{}$L_{0_{7\oplus\bar{1}}}$} & {\footnotesize{}$G_{2}^{4}(8),G_{2}^{5}(8)$} & \tabularnewline
\hline 
{\footnotesize{}4-qubit cluster state } & $G_{abcd}$ & {\footnotesize{}$G_{2}$} & {\footnotesize{}$G_{2}^{1}(8),G_{2}^{2}(8),G_{2}^{4}(8),G_{2}^{5}(8)$}\tabularnewline
\hline 
{\footnotesize{}4-qubit $\Omega$ state } & $L_{0_{3\oplus\bar{1}}0_{3\oplus\bar{1}}}$ & {\footnotesize{}$G_{2}$} & {\footnotesize{}$G_{2}^{i}(8)\,:i\in\left\{ 1,\cdots11\right\} $}\tabularnewline
\hline 
{\footnotesize{}4-qubit $Q_{4}$ } & {\footnotesize{}$L_{0_{5\oplus\bar{3}}}$} & {\footnotesize{}$G_{2}^{6}(8),G_{2}^{7}(8)$} & {\footnotesize{}$G_{2}^{5}(8)$}\tabularnewline
\hline 
{\footnotesize{}$\frac{\left|0001\rangle+|0010\rangle+|0111\rangle+|1011\right\rangle }{2}$} & $L_{ab_{3}}$ &  & {\footnotesize{}$G_{2}^{8}(8),G_{2}^{9}(8)$}\tabularnewline
\hline 
{\footnotesize{}$\frac{\left|0000\rangle+|0111\right\rangle }{\sqrt{2}}$} & {\footnotesize{}$L_{0_{3\oplus\bar{1}}0_{3\oplus\bar{1}}}$} &  & {\footnotesize{} $G_{2}^{4}(8),G_{2}^{5}(8),G_{2}^{8}(8),G_{2}^{9}(8),G_{2}^{10}(8),G_{2}^{11}(8)$}\tabularnewline
\hline 
{\footnotesize{}5-qubit Brown state } & - & {\footnotesize{}$G_{3}^{1}(32),G_{3}^{2}(32),G_{3}^{4}(32),G_{3}^{5}(32),G_{3}^{7}(32),G_{3}^{8}(32)$} & \tabularnewline
\hline 
{\footnotesize{}5-qubit cluster state } & - & {\footnotesize{}$G_{3}^{4}(32),G_{3}^{5}(32),G_{3}^{7}(32),G_{3}^{8}(32)$} & \tabularnewline
\hline 
\end{tabular}\caption{\label{table_group operators}List of useful quantum states and corresponding
operators that may be used to implement protocols for AQD and QD.
Specifically, to implement a protocol of QD using a state mentioned
in the 1st column of $i$th row of this table, both the users should
use one of the groups listed in Column 3 or 4 of the same row. However,
to implement an AQD, using a state mentioned in the 1st column of
$i$th row of this table, one of the user (say Bob) would use a group
($G_{B}$) listed in Column 3 or 4 of the same row, as was done in
QD, and Alice (who is expected to communicate less) would use a\textcolor{red}{{}
}smaller group $G_{A}:G_{A}\otimes I_{2}^{j-l}<G_{B},$ where Bob
and Alice encode their message by using $j$ and $l$ qubit unitary
operations, respectively.}
\end{table}

Let us consider a particular example in which a 4-qubit cluster state
is used for the implementation of a scheme for QD/AQD between Alice
and Bob. To begin with, let us consider the symmetric case in which
both Alice and Bob wish to communicate 4 bits of classical information
to each other. In this particular case, 2 travel qubits will be required,
and both Alice and Bob have to encode their secrets using the operators
from the group $G_{2}$. In contrast, they may also decide to go for
an AQD scheme using the same quantum state and allowing Bob to encode
4 bits of classical information and Alice to encode half of that.
In this particular case, Bob would still encode using the operators
from $G_{2}$ on two qubits (as he did in symmetric case), but would
send only one qubit to Alice, who will be able to encode her 2 bits
of classical information using unitary operators from $G_{1}$ group
on the travel qubit and send it back to Bob. Note that $G_{1}\otimes\{I_{2}\}=\{I_{2}\otimes I_{2},\,X\otimes I_{2},\,iY\otimes I_{2},\,Z\otimes I_{2}\}<G_{2}$
(i.e, $G_{1}\otimes\{I_{2}\}$ is a subgroup of $G_{2}$. Subsequently,
Bob will measure the final state in the suitable basis and broadcast
the measurement outcome. One may argue that we can implement the AQD
scheme using conventional QD, too. In that case, we have to send two
travel qubits through the channel. That would increase the possibilities
of being affected by the channel noise and also would involve higher
quantum cost as far as the communication via the quantum channel is
concerned. It is evident that if Alice needs to encode less classical
information then AQD with lesser number of travel qubits is preferable
and sufficient. In Table \ref{tab:Assym-QD}, a list of groups of
operators capable of implementing the proposed scheme of QD and AQD
for different quantum states is presented. It is shown that there
exists a large number of alternative ways (combination of groups of
operators and quantum states) that may be used for the implementation
of the proposed scheme for AQD. 

\begin{center}
\begin{table}
\begin{centering}
\begin{tabular}{|>{\centering}p{1.45cm}|>{\centering}p{0.5cm}|c|c|>{\centering}p{0.9cm}|c|>{\centering}p{2.4cm}|>{\centering}p{0.9cm}|}
\hline 
Quantum state  & \multicolumn{1}{>{\centering}p{0.5cm}}{} & \multicolumn{1}{c}{AQD} & \multicolumn{1}{c}{} &  & \multicolumn{1}{c}{} & \multicolumn{1}{>{\centering}p{2.4cm}}{QD} & \tabularnewline
\cline{2-8} 
 & $N_{T}$ & Operation of B & Operation of A & c-bits (B:A) & $N_{T}$  & Operation of B or A & c-bits (B:A)\tabularnewline
\hline 
2-qubit Bell state  & 1 & $g_{i}\,:i\in\left\{ 1,2,3\right\} $ & $G_{1}$ & 1:2 & 1 & $G_{1}$ & 2:2\tabularnewline
 & 1 & $G_{1}$ & $g_{i}\,:i\in\left\{ 1,2,3\right\} $ & 2:1 &  &  & \tabularnewline
\hline 
3-qubit $GHZ$  & 1 & $G_{2}^{i}(8)\,:i\in\left\{ 4,5\right\} $ & $g_{i}\,:i\in\left\{ 1,2,3\right\} $ & 3:1 & 2 & $G_{2}^{i}(8)\,:i\in\left\{ 4,5\right\} $ & 3:3\tabularnewline
 & 1 & $G_{2}^{i}(8)\,:i\in\left\{ 4,5\right\} $ & $G_{1}$ & 3:2 &  &  & \tabularnewline
\hline 
4-qubit cluster state  & 1 & $G_{2}$ & $g_{i}\,:i\in\left\{ 1,2,3\right\} $ & 4:1 & 2 & $G_{2}$ & 4:4\tabularnewline
and $\Omega$ state  & 1 & $G_{2}$ & $G_{1}$ & 4:2 &  &  & \tabularnewline
 & 2 & $G_{2}$ & $G_{2}^{i}(8)\,:i\in\left\{ 1,\cdots6\right\} $ & 4:3 &  &  & \tabularnewline
\hline 
5-qubit Brown state  & 1 & $G_{2}^{i}(8)\,:i\in\left\{ 1,2,4,5,7,8\right\} $ & $g_{i}\,:i\in\left\{ 1,2,3\right\} $ & 5:1 & 3 & $G_{2}^{i}(8)\,:i\in\left\{ 1,2,4,5,7,8\right\} $  & 5:5\tabularnewline
 & 1 & $G_{2}^{i}(8)\,:i\in\left\{ 1,2,4,5,7,8\right\} $ & $G_{1}$ & 5:2 &  &  & \tabularnewline
 & 2 & $G_{2}^{i}(8)\,:i\in\left\{ 1,2,4,5,7,8\right\} $ & $G_{2}^{i}(8)\,:i\in\left\{ 1,\cdots6\right\} $ & 5:3 &  &  & \tabularnewline
 & 2 & $G_{2}^{i}(8)\,:i\in\left\{ 1,2,4,5,7,8\right\} $ & $G_{2}$ & 5:4 &  &  & \tabularnewline
\hline 
\end{tabular}
\par\end{centering}

\caption{\label{tab:Assym-QD}Asymmetric quantum dialogue (AQD) and quantum
dialogue (QD) between Alice (A) and Bob (B). $N_{T}$ is the number
of travel qubits and $g_{1}=\{I_{2},X\}$, $g_{2}=\{I_{2},iY\}$,
and $g_{3}=\{I_{2},Z\}$. The notations used are elaborated in Appendix
A and are consistent with the previous table.}
\end{table}

\par\end{center}

\section{Robustness: Effect of noise on the AQD scheme \label{sec:Robustness:-Effect-of}}

To implement the AQD protocol described above, an entangled state
is to be used, part of which (travel qubits) will travel through the
channel and thus will get exposed to the environment, whereas the
other qubits (home qubits) would remain with one of the users. In
what follows, it will be assumed that home qubits will not be affected
by noise \cite{crypt-switch}. Keeping this in mind, in this section,
we aim to study the effect of widely used noise models: Amplitude
damping (AD) and Phase damping (PD) on the travel qubits in the AQD
protocol. We also compare the fidelity of the quantum state at the
end of Alice's and Bob's encoding in noisy and ideal scenarios obtained
for AQD in a particular type of noise channel, with that of the fidelity
obtained for QD implemented with the same conditions. 

Before we discuss the effect of noise on the proposed AQD scheme,
we will briefly introduce the strategy adopted here to quantify this.
If Bob prepares an $n$-qubit pure entangled state $\rho_{{\rm initial}}=\left[|\psi\rangle\langle\psi|\right]_{h+t}$
and keeps the home qubits ($h)$ with himself after sending travel
qubits ($t$) to Alice, where $t=n-h$, thus the noise acts on the
travel qubits only. The transformed quantum state in the presence
of noise can be written as 
\begin{equation}
\rho_{k}=\sum_{i_{j}}I_{2}^{\otimes h}\otimes E_{i_{1}}^{k}\otimes\cdots E_{i_{j}}^{k}\cdots\otimes E_{i_{t}}^{k}\rho_{{\rm initial}}\left(I_{2}^{\otimes h}\otimes E_{i_{1}}^{k}\otimes\cdots E_{i_{j}}^{k}\cdots\otimes E_{i_{t}}^{k}\right)^{\dagger},\label{eq:noise-effected-density-matrix-1}
\end{equation}
where $E_{i_{j}}^{k}$ are the suitable Kraus operators for AD or
PD channels, which will be discussed in detail in the following subsections. 

The effect of noise can be quantitatively obtained using a distance
based measure fidelity between the quantum state $\rho_{k}$ obtained
in the presence of noisy channels with the one in ideal condition.
Suppose the quantum state in the noisy channel after the encoding
of Alice and Bob is $\rho_{k}^{\prime}$, while it was expected to
be $|\psi^{\prime}\rangle$ in the absence of noise. In this case,
the fidelity is 
\begin{equation}
F_{k}=\langle\psi^{\prime}|\rho_{k}^{\prime}|\psi^{\prime}\rangle,\label{eq:fidelity}
\end{equation}
which is the square of the conventional fidelity expression and has
also been used in the past.

It would be worth mentioning here that we have obtained average fidelity
considering all possible encoding of Alice and Bob. Further, we have
considered here a special case with 4-qubit cluster state as initial
state, which will be utilized as quantum channel. AQD scheme gives
a certain advantage in the implementation of the scheme as it will
be less affected due to noise than the QD scheme. This is discussed
in following subsections.

\subsection{Effect of amplitude damping (AD) noise}

The effect of AD noise in the quantum state $\rho_{{\rm initial}}$
is characterized by the following Kraus operators \cite{nielsen}:
\begin{equation}
E_{0}^{A}=|0\rangle\langle0|+\sqrt{1-\eta_{A}}|1\rangle\langle1|,\,\,\,\,\,\,\,\,\,\,\,\,\,\,\,E_{1}^{A}=\sqrt{\eta_{A}}|0\rangle\langle1|,\label{eq:Krauss-amp-damping}
\end{equation}
where $\eta_{A}$ ($0\leq\eta_{A}\leq1$) is the decoherence rate
and describes the probability of error due to AD channel. 

The fidelity expressions are obtained for QD/AQD (with 2 travel qubits)
and AQD (with 1 travel qubit) as 
\begin{equation}
F_{AD}^{{\rm QD/AQD}}=\frac{1}{8}\left(\eta^{4}-4\eta^{3}+12\eta^{2}-16\eta+8\right)\label{eq:Fad-sy}
\end{equation}
and
\begin{equation}
F_{AD}^{{\rm AQD}}=\frac{1}{4}(\eta-2)^{2},\label{eq:Fad-asy}
\end{equation}
respectively. We have observed that fidelity expression depends on
the number of travel qubits (which depends on the amount of classical
information Alice wants to encode) and it does not depend on Bob's
encoding. Specifically, it does not depend on how much classical information
has been encoded by Bob, and how many qubits have been used for Bob's
encoding. This is not surprising as we have considered that the home
qubits don't get affected by the channel noise. Thus, as long as we
are considering that only travel qubits get affected by the channel
noise, our strategy for designing new schemes should be such that
a proposed scheme would utilize a minimum number of travel qubits.
More precisely, the number of travel qubits should be equal to the
minimum number of qubits that Alice requires to encode her message.
For example, one travel qubit would be sufficient as long as Alice's
message is up to 2 bits. Further, the presence of quadratic terms
in the fidelity expression of the encoded quantum state in AQD (with
1 travel qubit) are the signature of to and fro travel of the single
qubit between Bob and Alice. Similarly, the quartic terms in the fidelity
expression for QD/AQD with 2 travel qubits is the signature of to
and fro travel of qubits between Bob and Alice. 

\begin{figure}
\begin{centering}
\includegraphics[scale=0.7]{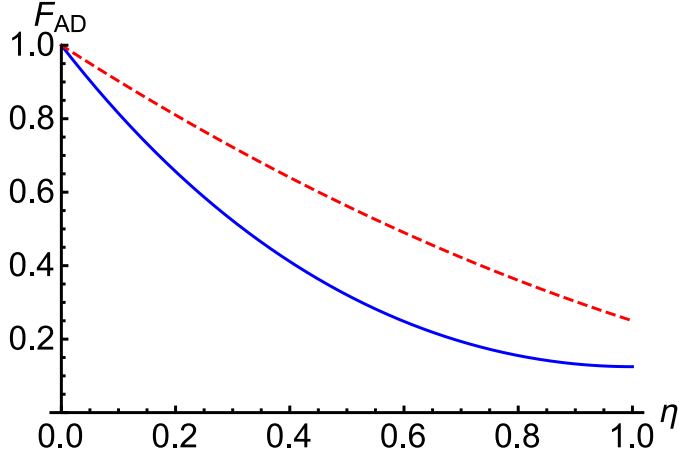}
\par\end{centering}

\caption{\label{fig:Effect-of-amplitutde}(Color online) The effect of AD noise
on the asymmetric QD depends on the number of travel qubits. Here,
4-qubit cluster state is used as quantum channel. The smooth (blue)
and dashed (red) lines correspond to 2 and 1 travel qubits, respectively.}
\end{figure}

To study the effect of AD channels, we have shown the variation of
both the fidelity expressions in Fig. \ref{fig:Effect-of-amplitutde},
which illustrates that the QD scheme is more affected due to noise.
Quantitatively, it can be seen that the obtained fidelity of the quantum
state carrying information in the AQD scheme is always greater than
that of the symmetric one.

\subsection{Effect of phase damping (PD) noise}

The effect of phase damping on the travel qubit is characterized by
the following Kraus operators \cite{nielsen} 
\begin{equation}
\begin{array}{ccc}
E_{0}^{P} & = & \sqrt{1-\eta_{P}}\otimes I,\\
E_{1}^{P} & = & \sqrt{\eta_{P}}|0\rangle\langle0|,\\
E_{2}^{P} & = & \sqrt{\eta_{P}}|1\rangle\langle1|,
\end{array}\label{eq:Krauss-phase-damping}
\end{equation}
where $\eta_{P}$ ($0\leq\eta_{P}\leq1$) is the decoherence rate
for the phase damping.

Similar to AD noise case, the fidelity expression for the symmetric
or asymmetric QD schemes with 2 travel qubits is 
\begin{equation}
F_{PD}^{{\rm sy}}=\frac{1}{2}\left((\eta-1)^{4}+1\right),\label{eq:Fpd-sy}
\end{equation}
whereas 
\begin{equation}
F_{PD}^{{\rm asy}}=\frac{1}{2}\left(\eta^{2}-2\eta+2\right)\label{eq:Fpd-asy}
\end{equation}
is obtained with 1 travel qubit, when the travel qubits are passing
through a PD channel. Therefore, the presence of quadratic (quartic)
terms are signature of forward and backward communication of 1 (2)
travel qubit(s).

\begin{figure}
\centering{}\includegraphics[scale=0.7]{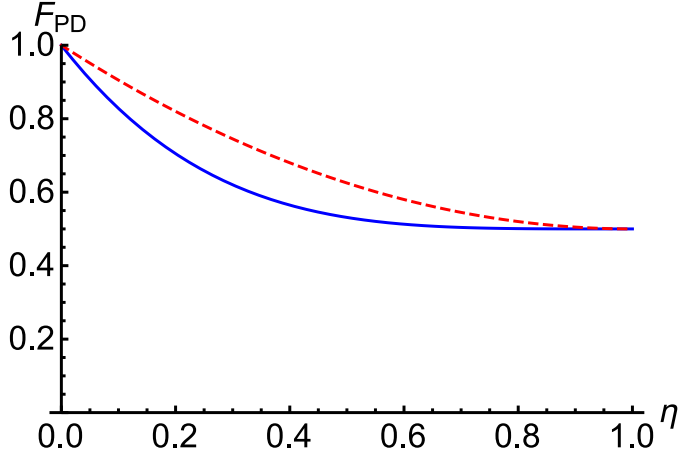}\caption{\label{fig:Effect-of-noise}(Color online) The dependence of the fidelity
of the information encoded quantum state of the AQD schemes on the
number of travel qubits is shown, when subjected to PD noise using
4-qubit cluster state as quantum channel. The smooth (blue) and dashed
(red) lines correspond to 2 and 1 travel qubits, respectively. }
\end{figure}

The variation of both the fidelity expressions in Fig. \ref{fig:Effect-of-noise}
further establishes that the AQD scheme is less affected and is preferable
over the symmetric one. Though, for very large values of decoherence
rate, the obtained fidelity of the information encoded quantum state
in both the cases become the same, otherwise the fidelity for the
AQD scheme is always greater than that of the symmetric one.

\section{Leakage and efficiency of the proposed scheme \label{sec:Leakage}}

Leakage is inherent in the Ba-An-type QD (for a discussion on this
see \cite{book,qd}), and the same limitation is applicable to the
proposed AQD protocol, too. In brief, leakage can be thought of as
the difference between the total information sent by both the legitimate
users and the minimum information required by Eve to extract that
information. Specifically, Eve requires the encoding information of
at least one of the users to extract the information of the other
user. For instance, in Ba An QD scheme, the total amount of encoded
classical information is 4 bits and minimum encoding of a party is
2 bits (i.e., Eve requires 2 bits of minimum information) resulting
in leakage of 2 bits. In case of AQD, the minimum requirement for
Eve becomes less than QD hence resulting in increase of leakage. However,
the leakage can be avoided if the initial state is unknown to the
users. In Ba An's original scheme \cite{ba-an}, it is explicitly
mentioned (in connection with the choice of initial state) that ``The
choice may be random or in some secret fashion unknown to Eve''.
The choice signifies which particular Bell state is selected among
the four Bell states. Here, we have taken this approach for AQD and
we conjecture that if we can send the initial state to the users by
QSDC (secret fashion) then the leakage can be nullified. 

We may now look at the qubit efficiency of the scheme with and without
QSDC. To do so, we use following measure of qubit efficiency \cite{defn of qubit efficiency} 

\begin{equation}
\eta=\frac{c}{q+b},\label{eq:efficiency}
\end{equation}
where $c$ denotes the total number of transmitted classical bits
(message bits), $q=Q+2t$ denotes the total number of qubits used
with $Q$-qubit entangled state as quantum channel and $t$ travel
qubits (corresponding to decoy qubits), $b$ is the number of classical
bits exchanged for decoding of the message (classical communication
used for checking of eavesdropping is not counted). If we consider
the example of Bell-state-based Ba-An-type QD protocol, then $c=4,$
$q=2+2=4$ and $b=2$, and thus efficiency is calculated to be 67\%.
Now, if we use a Bell state for QSDC of the information regarding
initial state, then total $q=8$ and thus efficiency is less. However,
we know that we need to send the initial state, just once. Thus, even
if we have to send $n$ c-bits via the main part of the QD scheme,
we still need to communicate 2 bits of classical information regarding
the initial state through a QSDC scheme. This would require 4 extra
qubits (2 qubits for channel and 2 qubits for eavesdropping checking)
and thus in the $n\rightarrow\infty$ limit the efficiency would become
the same as that in the QD scheme without QSDC. This is so because
in this particular case, $\eta=\underset{n\rightarrow\infty}{\lim}\frac{4n}{\left(2n+2n\right)+4+2n}=\underset{n\rightarrow\infty}{\lim}\frac{4n}{6(n+\frac{2}{3})}\approx67\%.$ 

As we have discussed the specific case of 4-qubit cluster state as
a quantum channel for the discussion of the effect of noise in the
previous section, it would be relevant to calculate and discuss the
efficiency for the same. Specifically, in the QD protocol with 4-qubit
cluster state, the efficiency would be $\eta=\frac{8}{\left(4+4\right)+4}=67\%$.
However, in AQD protocol, we decrease the value of $c$ and $t$ by
2 and 1 respectively to obtain qubit efficiency $\eta=60\%$. It may
appear that the efficiency will always decrease with AQD protocols
when compared with their QD counterparts. A contrary example would
be of GHZ state, where the efficiency can be calculated to be 60\%
and 62.5\% for QD and AQD protocols, respectively. It proves that
if we have a multipartite state with odd number of qubits then efficiency
can be increased in AQD protocols. Further, it has been already established
that using QSDC for sending the initial state information, the efficiency
equal to that of the protocols without using QSDC can be obtained
for a large number of copies of the quantum channel.

\section{Conclusion \label{sec:Conclusion}}

We have designed a protocol for AQD which is different from the standard
protocols of QD in the sense that the communication powers of the
users are different. Interestingly, the task is performed without
violating the requirements of QD. In the process of design and analysis
of this protocol, we have obtained several interesting results. For
example, we obtained a large number of new alternatives (in terms
of groups of operators and corresponding quantum states), which can
be used to implement QD (cf. last column of \ref{table_group operators}).
Secondly, we obtained a group theoretic structure of the operators
that may be used to realize the proposed scheme for AQD. The analysis
performed on the proposed scheme has also revealed that the proposed
AQD involves more leakage in comparison to its QD counterpart. However,
the leakage can be completely circumvented by including a QSDC scheme
for sharing the information about the initial state prepared by Bob.
Further, as the number of travel qubits is reduced in AQD (in comparison
to an equivalent QD scheme), the effect of various noise models is
also reduced. Further, using GHZ state as an example, it has been
shown that the qubit efficiency of the proposed AQD is higher than
the corresponding scheme of QD. The present AQD scheme can be easily
extended to design a controlled AQD scheme, where a controller (Charlie)
prepares the quantum state, which is followed by his announcement
of initial state after the HJRSP scheme (except the unitary operations
of Bob, whhich would depend on the initial state prepared by Charlie)
has been faithfully implemented by Alice and Bob. Charlie can also
send the information regarding initial state using QSDC scheme to
both the legitimate users, Alice and Bob. Thus, in brief, present
work introduces the concept of AQD, which is shown to be much beneficial
(in both noiseless and noisy environments) than a QD in a situation
where the users are not required to communicate equal amount of information,
and the proposed scheme can also be extended to develop a relevant
scheme for controlled-quantum communication. 

\textbf{Acknowledgment:} AB acknowledges support from the Council
of Scientific and Industrial Research, Government of India (Scientists'
Pool Scheme). CS thanks Japan Society for the Promotion of Science
(JSPS), Grant-in-Aid for JSPS Fellows no. 15F15015. She also thanks
IISER Kolkata for the hospitality provided during the initial phase
of the work. KT and AP thank Defense Research \& Development Organization
(DRDO), India for the support provided through the project number
ERIP/ER/1403163/M/01/1603.

\section*{Appendix A: Modified Pauli groups and the notation used \label{sec:Appendix-A:-Modified}}

Modified Pauli groups and the notations used to denote them in the
present work was introduced earlier in \cite{qd,QKA-cs}. Here, for
consistency, we briefly summarize the definition and the notation
used.

\setcounter{equation}{0} \renewcommand{\theequation}{A.\arabic{equation}}  

It is easy to verify that the set of Pauli operators $\{I_{2},\,\sigma_{x},\,i\sigma_{y},\,\sigma_{z}\}$
forms a group under multiplication $G_{1}^{\prime}=\left\{ \pm I_{2},\pm iI_{2},\pm\sigma_{x},\pm i\sigma_{x},\pm\sigma_{y},\pm i\sigma_{y},\pm\sigma_{z},\pm i\sigma_{z}\right\} $
(cf. Section 10.5.1 of \cite{nielsen}). The closure property of $G_{1}^{\prime}$
is satisfied under normal matrix multiplication because of the inclusion
of $\pm1$ and $\pm i$ . However, if any of the operators $\sigma_{i},\,-\sigma_{i},\,i\sigma_{i}$
or $-i\sigma_{i}$ operates on a quantum state, the effect would be
the same. Keeping this in mind, if global phase is ignored from the
product of matrices (which is consistent with quantum mechanics),
we obtain a modified Pauli group $G_{1}=\{I_{2},\,\sigma_{x},\,i\sigma_{y},\,\sigma_{z}\}=\{I_{2},\,X,\,iY,\,Z\}$.
Clearly, under the above defined multiplication rule, $G_{1}$ is
an Abelian group of order 4 and its generators are $\left\langle X,Z\right\rangle ,\,\left\langle X,iY\right\rangle \,{\rm and\,}\left\langle iY,Z\right\rangle $.
Similarly, we may define the modified generalized Pauli group $G_{n}=G_{1}^{\otimes n}$
as a group of order $2^{2^{n}}=4^{n}$ and whose elements are all
$n$-fold tensor products of Pauli matrices \cite{qd}. For example,
\begin{equation}
\begin{array}{lcl}
G_{2} & = & G_{1}\otimes G_{1}=\{I_{2},\,X,\,iY,\,Z\}\otimes\{I_{2},\,X,\,iY,\,Z\}\\
 & = & \left\{ I_{2}\otimes I_{2},\,I_{2}\otimes X,\,I_{2}\otimes iY,\,I_{2}\otimes Z,\,X\otimes I_{2},\,X\otimes X,\right.\\
 &  & X\otimes iY,\,X\otimes Z,\,iY\otimes I_{2},\,iY\otimes X,\,iY\otimes iY,\\
 &  & \left.iY\otimes Z,Z\otimes I_{2},\,Z\otimes X,\,Z\otimes iY,\,Z\otimes Z\right\} .
\end{array}\label{eq:pustak2}
\end{equation}
In Ref. \cite{qd}, it was discussed in detail, how to construct subgroups
of $G_{2}$. Here, we list 11 subgroups of $G_{2}$, which are used
in this paper (each is of order 8): 
\begin{equation}
\begin{array}{lcl}
G_{2}^{1}(8) & = & \left\{ I_{2}\otimes I_{2},\,X\otimes I_{2},\,iY\otimes I_{2},\,Z\otimes I_{2},\,I_{2}\otimes X,\,X\otimes X,\,iY\otimes X,\,Z\otimes X\right\} ,\\
G_{2}^{2}(8) & = & \left\{ I_{2}\otimes I_{2},\,X\otimes I_{2},\,iY\otimes I_{2},\,Z\otimes I_{2},\,I_{2}\otimes iY,\,X\otimes iY,\,iY\otimes iY,\,Z\otimes iY\right\} ,\\
G_{2}^{3}(8) & = & \left\{ I_{2}\otimes I_{2},\,X\otimes I_{2},\,iY\otimes I_{2},\,Z\otimes I_{2},\,I_{2}\otimes Z,\,X\otimes Z,\,iY\otimes Z,\,Z\otimes Z\right\} ,\\
G_{2}^{4}(8) & = & \left\{ I_{2}\otimes I_{2},\,I_{2}\otimes X,\,I_{2}\otimes iY,\,I_{2}\otimes Z,\,X\otimes I_{2},\,X\otimes X,\,X\otimes iY,\,X\otimes Z\right\} \\
G_{2}^{5}(8) & = & \left\{ I_{2}\otimes I_{2},\,I_{2}\otimes X,\,I_{2}\otimes iY,\,I_{2}\otimes Z,\,iY\otimes I_{2},\,iY\otimes X,\,iY\otimes iY,\,iY\otimes Z\right\} \\
G_{2}^{6}(8) & = & \left\{ I_{2}\otimes I_{2},\,I_{2}\otimes X,\,I_{2}\otimes iY,\,I_{2}\otimes Z,\,Z\otimes I_{2},\,Z\otimes X,\,Z\otimes iY,\,Z\otimes Z\right\} \\
G_{2}^{7}(8) & = & \left\{ I_{2}\otimes I_{2},I_{2}\otimes Z,Z\otimes I_{2},Z\otimes Z,X\otimes X,iY\otimes X,X\otimes iY,iY\otimes iY\right\} ,\\
G_{2}^{8}(8) & = & \left\{ I_{2}\otimes I_{2},Z\otimes Z,X\otimes iY,iY\otimes X,I_{2}\otimes X,\,Z\otimes iY,\,iY\otimes I_{2},\,X\otimes Z\right\} ,\\
G_{2}^{9}(8) & = & \left\{ I_{2}\otimes I_{2},Z\otimes Z,X\otimes iY,iY\otimes X,X\otimes I_{2},iY\otimes Z,Z\otimes X,I_{2}\otimes iY\right\} ,\\
G_{2}^{10}(8) & = & \left\{ I_{2}\otimes I_{2},X\otimes I_{2},I_{2}\otimes X,X\otimes X,Z\otimes Z,iY\otimes Z,Z\otimes iY,iY\otimes iY\right\} ,\\
G_{2}^{11}(8) & = & \left\{ I_{2}\otimes I_{2},iY\otimes I_{2},I_{2}\otimes iY,iY\otimes iY,Z\otimes Z,Z\otimes X,X\otimes Z,X\otimes X\right\} ,
\end{array}\label{eq:pustak3}
\end{equation}
where $G_{n}^{j}(m)$ denotes $j^{th}$ subgroup of order $m<4^{n}$
of the group $G_{n}$ whose order is $4^{n}.$

\end{document}